# Demonstration of Neutron Radiation-Induced Nucleation of Supercooled Water

Matthew Szydagis[1,*], Cecilia Levy[1,*], Yujia Huang[1], Alvine C. Kamaha[1], Corwin C. Knight[1], Gregory R. C. Rischbieter[1], and Peter W. Wilson[2,3*]

[1] The University at Albany, SUNY
Department of Physics
1400 Washington Avenue
Albany, NY, USA 12222-0100

[2] School of Biological, Earth, and Environmental Sciences
University of New South Wales, NSW 2052, Australia

[3] The King's Institute, The King's School
Parramatta, NSW 2151, Australia

*co-corresponding authors: pwilson@kings.edu.au, mszydagis@albany.edu, clevy@albany.edu

## Abstract

We present here direct evidence for neutrons causing nucleation of supercooled water. Highly purified water (20 nm filtration) is cooled to well below freezing (as low as -20 °C) with a radioactive calibration source of neutrons / gamma-rays either present or removed during each of many control cooling runs for the same volume of water. When it is primarily neutrons irradiating the sample bulk, the non-equilibrium freezing point (also known as the "supercooling point") is, on average, +0.7 °C warmer than the control equivalent, with a statistical significance of greater than 5 sigma, with systematic uncertainty included. This effect is not observed with water in the presence of gamma-rays instead of neutrons. While these neutrons should have theoretically had sufficient energy to mount the energy barrier, corroborating our results, their raising of supercooling temperature has never been reported experimentally to the best of our knowledge. The potential to use deeply supercooled solutions, not only water, as metastable detectors for radiation and perhaps dark matter or neutrino physics presents now a new avenue for exploration.

## INTRODUCTION

The heterogeneous nucleation of supercooled water and aqueous solutions continues to be the subject of study across diverse fields ranging from atmospheric physics[1] to cryobiology[2]. Nucleating particles and their efficiencies are well known for mineral salts[3], for crystals such as silver iodide[4], and for biological molecules such as the ice-nucleating proteins associated with, for instance, the bacterium *pseudomonas syringae*[5]. However, the effects of neutrons and other radiation types on metastable water have not been studied in detail. While *a priori* expected, the question of whether such particles can directly cause some water molecules (H and/or O atoms) to gain sufficient kinetic energy to mount the energy barrier (if we are considering classical nucleation theory[6]) and therefore cause an irreversible nucleating event, has not been experimentally tested, to the best of our knowledge.

This lack of clarity around the effects of irradiation upon supercooled water, especially for neutrons, is partly due to the inherent stochastic nature of nucleation and the fact that many freeze / thaw cycles are required on the same liquid sample. This repetition is required in order to have meaningful statistics on the nucleation of that sample, within that container, both with / without the potential nucleator (or anti-nucleator) in question.

Perhaps the most well-accepted approach to analyzing nucleation statistics to date is that by Haymet and colleagues[7,8], who termed the experimental arrangement "ALTA," which stands for Automatic Lag Time Apparatus, and who have looked at, as examples: the addition of crystals of AgI[9], stirring[10], coating the container with a hydrophobic surface[11], and many other alterations which may affect the nucleation of a sample of water (and other aqueous solutions) in a given container.

To the best of our knowledge, the only references for radiation-induced nucleation of supercooled water are early works by Varshneya[12,13,14]. However, familiarity with plotting survival curves (and skew-Gauss fits) of the nucleation events on a series of runs (in the way of ALTA) allows one to identify where the mean supercooling temperature (or so-called $T_{50}$) lies, and whether it is demonstrably different from a control when in the presence of, for instance, additives or of neutron sources, as described here. Perhaps the best demonstration of the power of the survival curve is the "Manhattan" of Heneghan et al., or the bubble or no-bubble plot (Figure 4 in ref [15]).

## EXPERIMENTAL SETUP

The concept is shown in Figure 1, and the actual experimental setup is in Figure 2. A fused quartz tube, ultrasonically cleaned within an ISO3 cleanroom, was prepared with 22 ± 1 grams of water, then fully submerged in a thermoregulation ethanol bath set over vibration-dampening pads and instrumented with three thermocouples for the recording of exothermic increases on freezing. These were attached near the top, the middle (water line), and the bottom (hemisphere). Supplementing these 3 thermometers, a fourth one, whose variation had no discernible effect, recorded room temperature. LEDs provided illumination from below while a borescope provided images.

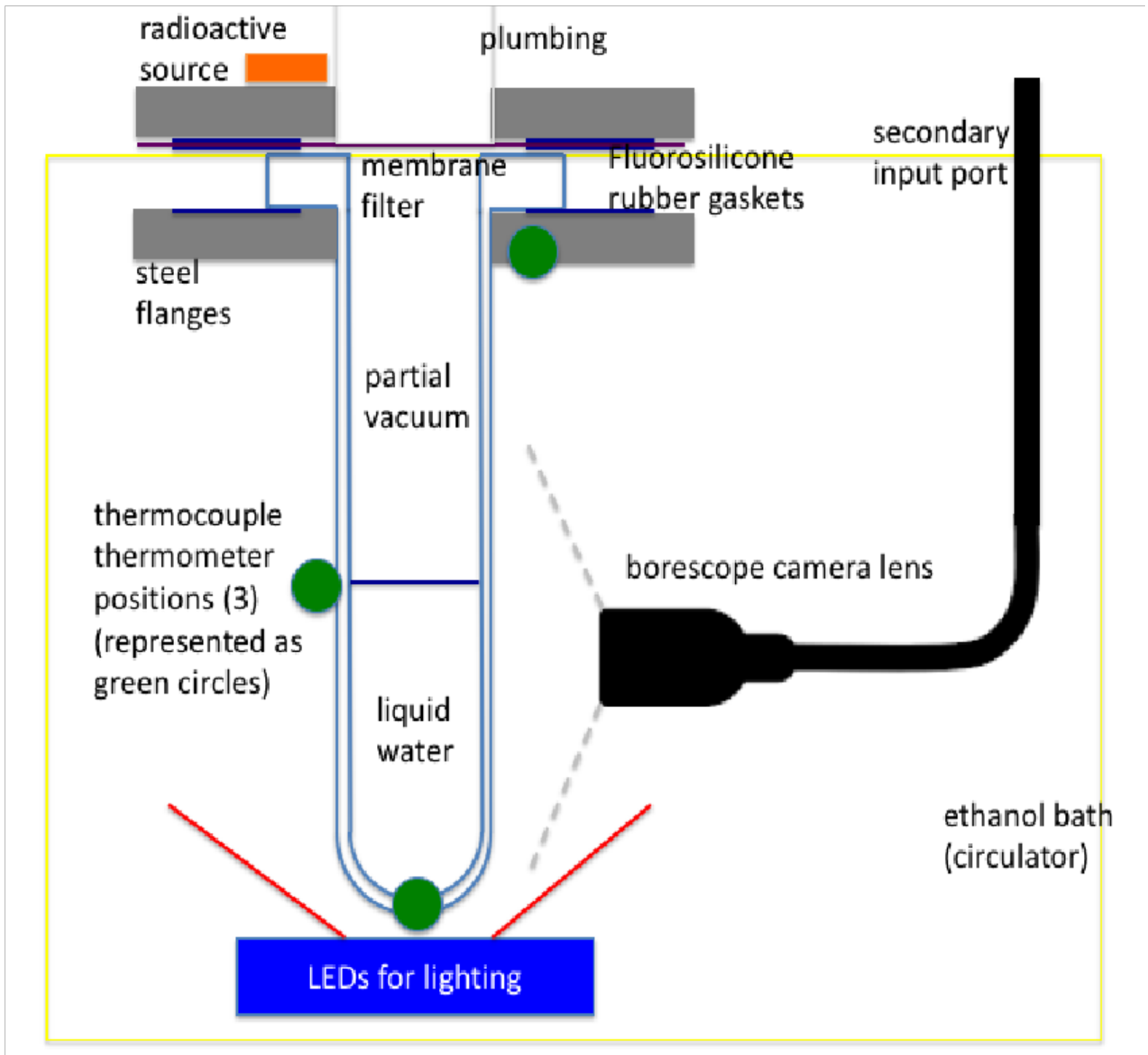

FIG. 1. Diagram of setup, the core of which was one 10-cm-long, cylindrical quartz tube (left) with inner diameter of 3.55 cm, outer 4.20 cm, and hemispherical bottom.

Purity of the water is of importance to minimize the temperature when heterogeneous nucleation occurs, where eliminating potential nucleation sites such as dirt and other particulates leaves only the wall of the container as the source of nucleation. It should be noted that heterogeneous nucleation will always occur upon a two-dimensional surface, rather than within three dimensions, which would be homogeneous and occurs at ~ -40 °C[16,17], except perhaps in the case where an incident neutron causes a water molecule to mount the energy barrier (an idea developed in this work).

Deionized water was distilled through a 20-nm porous filter membrane into a quartz tube evacuated via an oil-less pump. Note this small-diameter filter required a very long process (months) in order to force the water through. The water was cooled in an ethanol bath until it reached a supercooled state, at which point an incident particle could theoretically cause the phase transition.

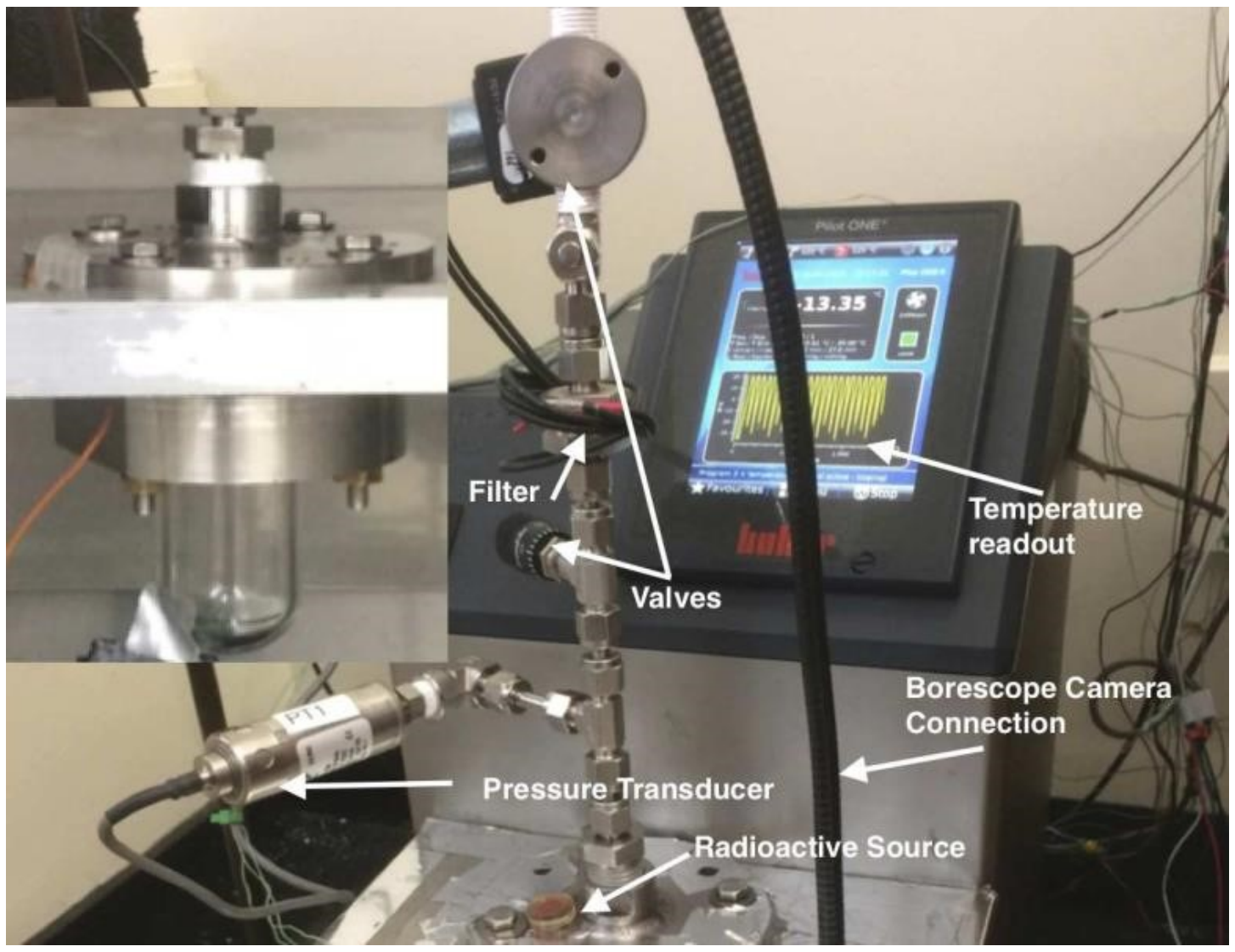

FIG. 2. Photo of setup during operation. Left inset is quartz tube containing 22 mL of purified water.

The water was continuously cooled at a mean rate of 1.90±0.05 °C/min, which likely caused temperature non-uniformity which slowed nucleations. The data were taken both with and without a fission neutron source, i.e., $^{252}$Cf, which produces a wide spectrum of n's with typical energies $O$(1-10) MeV[18]. Its overall activity was 1.0 $\mu$Ci, leading to ~3000 n / s[19]. Subsequent (one year later) data was also taken both with and without a gamma ray source of $^{137}$Cs. Data were taken in runs, with each run containing several data sets, alternating control and source data, to minimize systematics. A run always included at least one control data set, and only included data from 1 source at the same location (for instance, a run was alternation of control-Cf-control-Cf-control data sets). Each data set is either 2-day-long or 1d-long, and consists of several tens of individual events. An *event* was defined as a freezing of the liquid, as the freezing event is still rapid for deeply supercooled water, when compared to the time spent supercooled.

Approximately equal numbers of data sets with a source vs. none (control) were collected across day and night. Beginning at 20 ˚C, 1 cool-down took about 30 minutes (slope indicated in Figure 3). The thermo-regulator was set so that the climb to 20 ˚C was 30 minutes as well, to set up the next 1-hour-long cycle.

The initial data analysis was based on studying two parameters: the temperature $T_{min}$ at which the freezing event occurred, and the time $\Delta t_{active}$ that the water had spent in a supercooled state (given approximately linear cooling, these should be equivalent). More precisely, $\Delta t_{active}$ is defined as the time interval between -15.5 ˚C and when freezing occurs, with -15.5 ˚C selected as the higher temperature border above which no nucleation is observed (see Figure 4). This analysis was then refined into the more traditional S-curve style of ALTA, effectively summarizing all of the important data more concisely. The survival, or "S," curve shows unfrozen fraction as a function of temperature (black and yellow lines in Figure 4 left). The $T_{50}$'s are then the temperatures at which the sample had frozen 50% of the time.

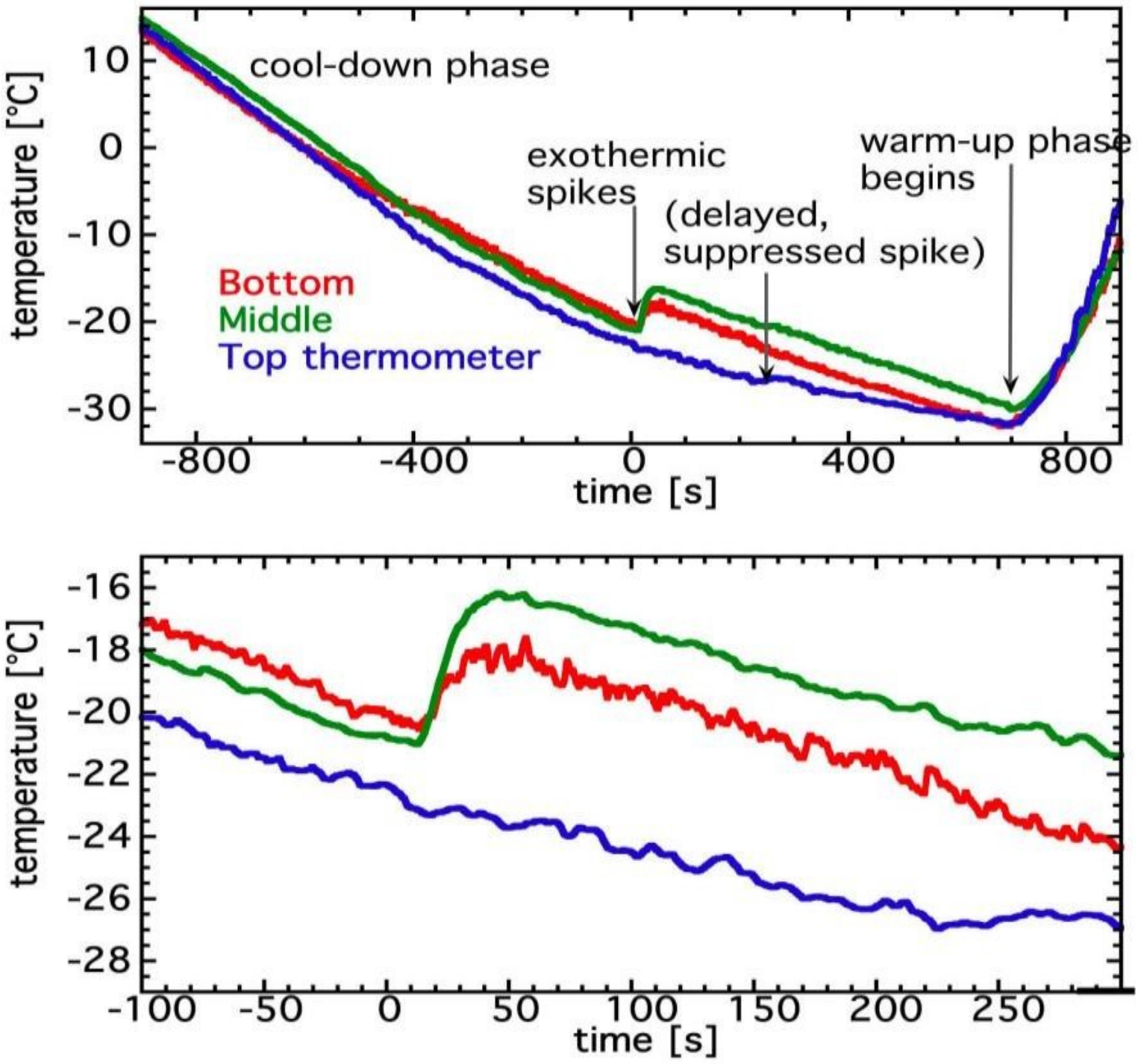

FIG. 3. *Top* The temperature profile for a typical event. *Bottom* Zoomed-in version near $t$ = 0. The exothermic rise at nucleation (and the subsequent freezing) can be clearly seen in the green and red traces from two different thermometers. The blue trace is from a thermometer placed some distance from the sample, at the top of the tube.

$\Delta t_{active}$ has the advantage of the moment of freeze being visually verifiable, but the disadvantage of still relying on $T$ for defining the start time. $T$'s were always measured outside the quartz to avoid nucleation, leading to a universal 2.5 ˚C offset at the cooling rate set, already added to all of the values reported here, control and source alike, increasing them compared to what was actually measured. This renormalizing offset was determined by viewing the plateau in $T$ in the top thermocouple during melting, consistently occurring at 2.5, not 0.0 ˚C (where, during melting, the offset had the opposite direction with the sample colder than measured, due to the thermal lag during heating). All three thermocouples were also cross-calibrated against the built-in RTD of the chiller. Lastly, the thermocouples reacted differently based on location. We minimize that last systematic in the choice of the top for defining minimum temperatures, for the purpose of a greater consistency amongst runs, as it was the only thermometer never requiring reattachment, thus making it most consistent despite it being furthest from the water and reacting the weakest/latest (Figure 3, blue). Camera images were used to determine and verify the correct event time, to determine the best time to use for pulling a recording of temperature from the top thermometer.

In order to mitigate any systematic effects of any inherent stochastic changes during the data-taking, like in the cooling rate, we compute the statistical significances of any source data with respect to control by considering only control sets from within the same one run, immediately preceding and/or following the source sets.

## RESULTS

When the $^{252}$Cf neutron source was present, water did not remain supercooled as long, on average, over multiple cool-downs. Accordingly, freezing occurred at warmer minimum temperatures $T_{min}$, as indicated by a +5.8$\sigma$ difference in raw median, and +5.2$\sigma$ difference in $T_{50}$ (the temperature at which the freezing had occurred 50% of the time across all data sets in the run) between $^{252}$Cf vs. control data sets. All $T_{50}$'s are derived from an error function erf (Gaussian integral) fit, most common in the field. The effect can be seen in Figure 4 left. At right, skew-Gaussian and normal Gaussian fits to the raw data and the unbinned population mean are all explored as cross-checks. For control, a skew fit is superior, consistent with past supercooled water data[8], but standing in contrast with the neutron data. Statistical errors are based upon combining all data points of one type from the same run and calculating the population standard deviation divided by the square root of the number of events included from all data sets within the one same run.

Systematic error in temperature is conservatively determined by the deviation in the median $T_{min}$'s across a calendar year of >800 control events, with individual data sets defined by 24 or 48-hr long collections of events. This was computed by first subtracting the medians of the control values from all of the control values, by run, to centrally renormalize all of the data. This is followed by computing the standard error, defined as for statistical error. Choice of median over, for example, skew-Gaussian centroid, is motivated by it being less biased, by not being subject to the choice of binning, nor assumptions on shape. Medians are also less affected by extreme outliers that could dramatically shift arithmetic means. No data are cut.

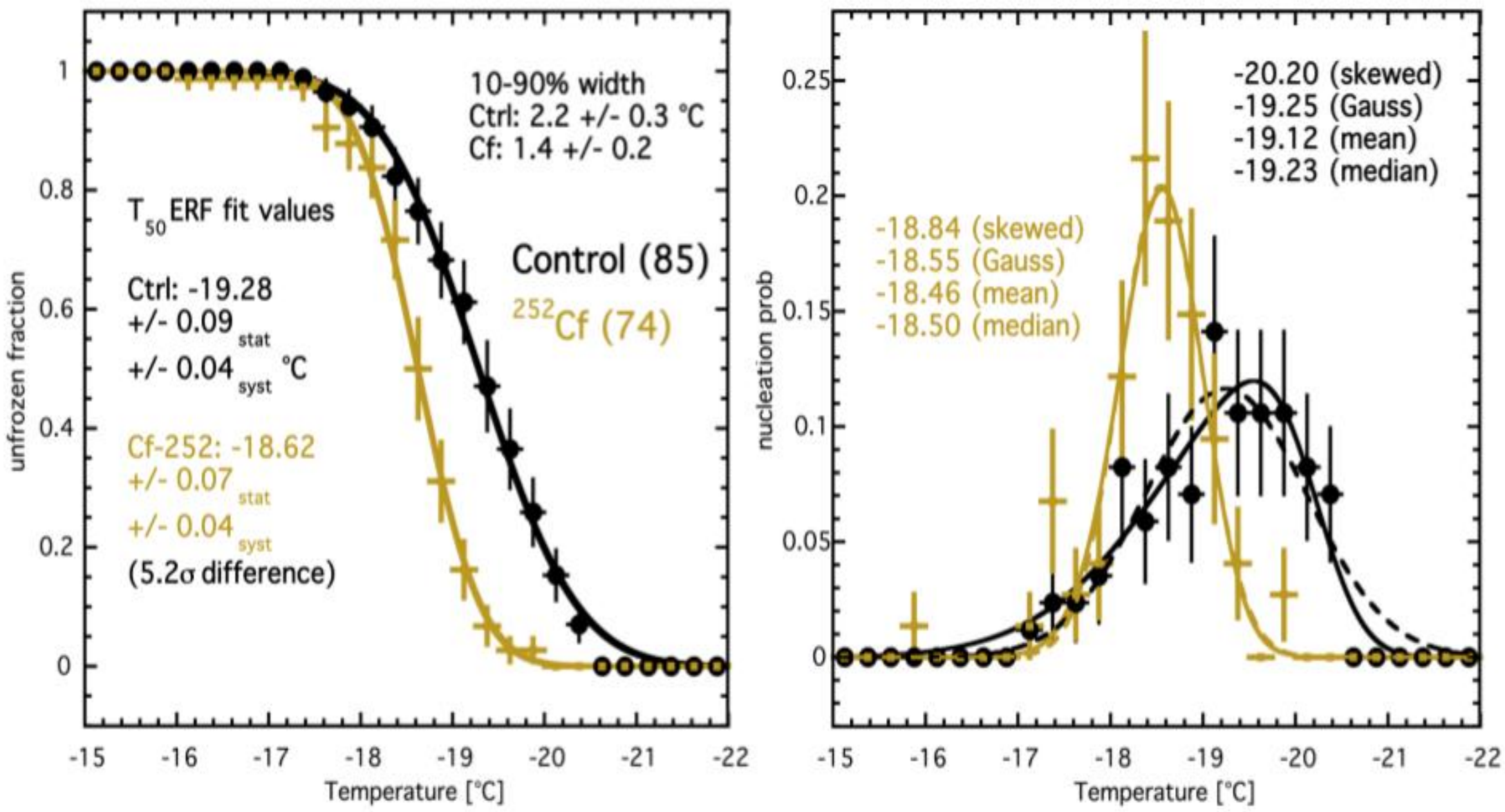

Figure 4. (Left) All $^{252}$Cf data (74 events) in yellow, and control data from the same period of time (85 events) in black. For control, only the data immediately before, after, and between Cf data is used. X-errors are bin widths (0.25 °C) while y-errors are statistical only (Poisson). $T_{50}$'s (-19.28 and -18.62) quoted are the T's at which the sample had frozen 50% of the time, based upon the smooth error-function fits, solid lines. The FWHM's of skew fits are comparable to the 10-90% widths of the survival curves fit by erf's. (Right) Skew solid and Gaussian dashed, each different for control but very similar for Cf, with Cf's skewness parameter $\alpha$ = 1.0 +/– 0.9, thus consistent with zero (for control, $\alpha$ = 3.4 +/– 1.1).

In addition to $T_{50}$, the 10-90% widths of the survival curves are also of interest. Previously, Wilson and Haymet[20] have shown that with sufficient runs on the same sample the 10-90 width seems to converge to 0.72 °C. Our control exhibits 2.2 °C, suggesting more than one nucleation site within the glass container and water volume, and/or a nucleation site that changes over time. Both hypotheses are valid according to a visual analysis of data, both control and source (Figure 5). The irradiated runs show a narrower width, 1.4 °C, and as this nucleation mechanism is by direct momentum transfer likely to the water in the bulk (rather than directly to water momentarily at the glass surface) and to different water molecules by event, the difference and the overall effect certainly warrant further investigation (but beyond the scope of this paper). The narrowing may be due to more events at the same/similar locations, close to the source.

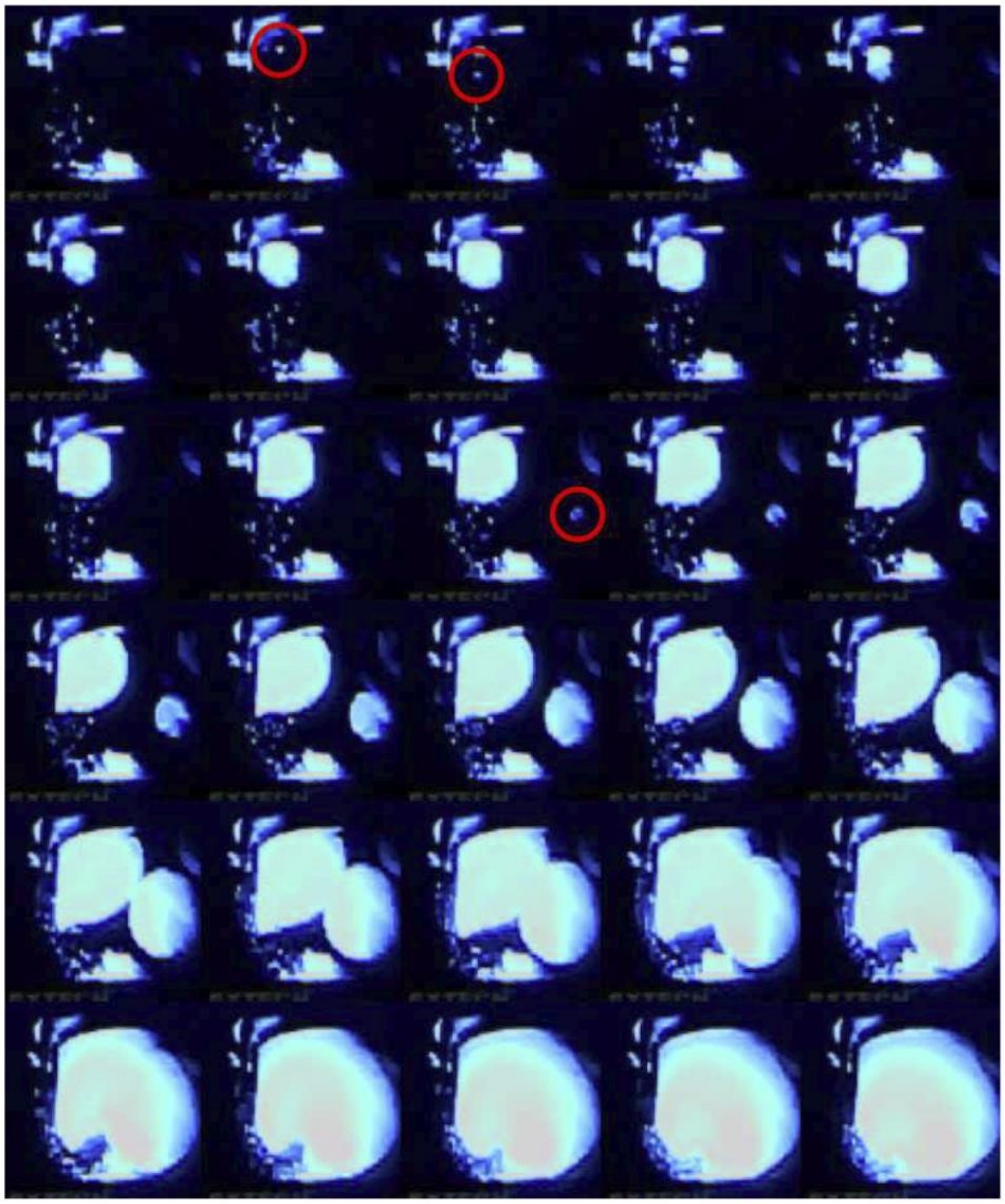

Figure 5. An example of one possible triple-nucleation event in neutron data. Red circles indicate the first frames in which a nucleation site appears. The first two snowballs merge rapidly, but a 3rd appears much later, implying that it is from a distinct neutron. Unlike in a bubble chamber there is no pressure increase activated post trigger so any unfrozen volume remains active during an ongoing event. For brevity, only every 3rd frame is pictured, that is, every 150 ms. Due to the large volume, thermal gradients were likely present in both control and source data, causing this slow freezing, visibly taking seconds, but having no impact on the actual result of the experiment.

Space constraints in the thermo-regulating chiller forced us to have only one camera, thus making three-dimensional position reconstruction impossible. As a result, any quantitative conclusions from images is forced to wait until future work when it should be possible to determine surface versus bulk events more readily and thus heterogeneous versus homogeneous nucleation, using a minimum of two cameras offset in angle, and any differences between control and source and/or between single- and multiple-scattering. Nevertheless, as demonstrated in Figure 5 as a sample, our proof-of-concept preliminary work presented here is sufficient for concluding that multiple sites were indeed active, both spatially and temporally.

Despite a formal image analysis lacking at this time additional confirmation of the primary result was still possible, via the breakdown of the Cf run into individual data sets, in Figure 6.

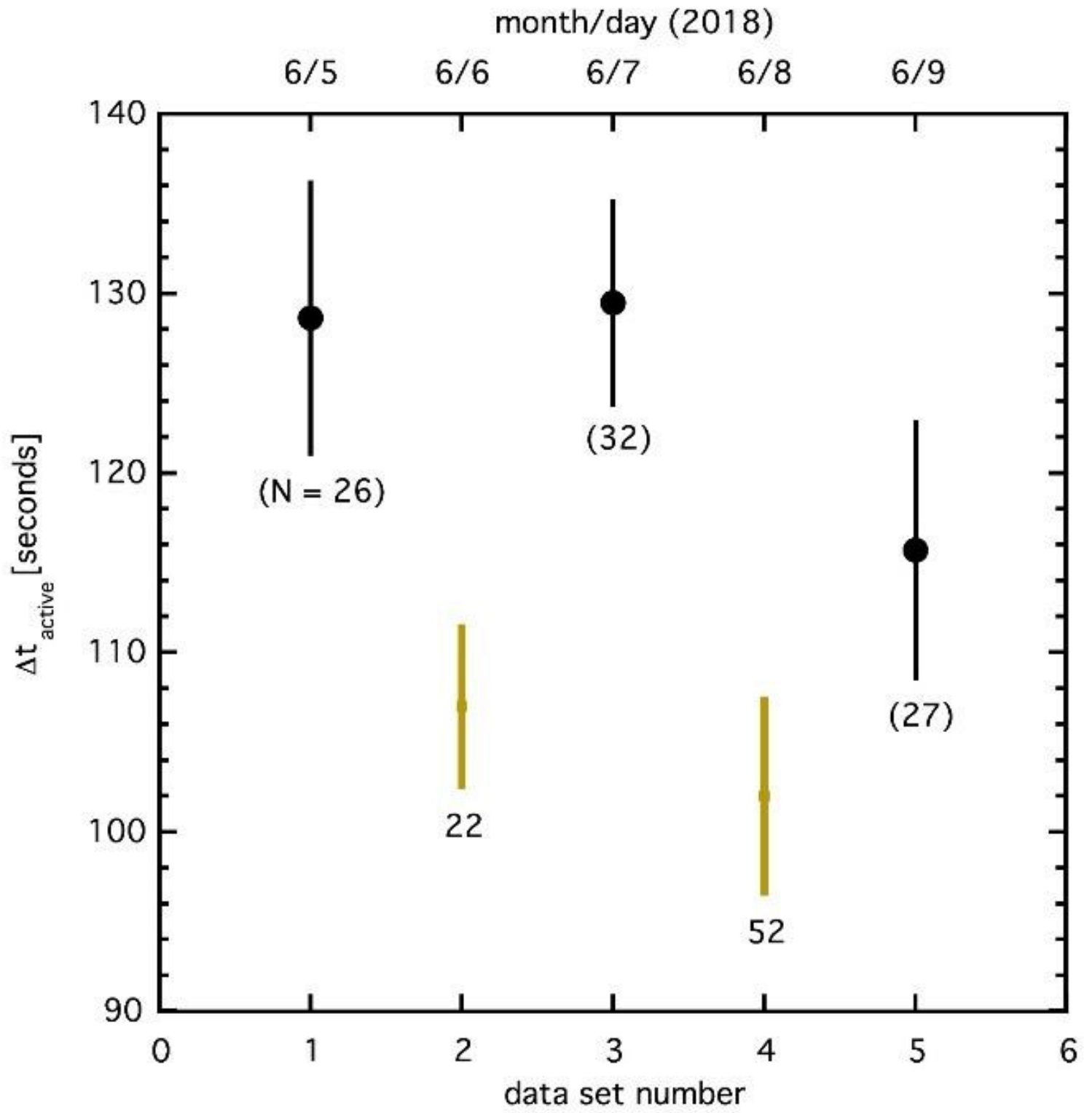

Figure 6. Control data sets for the five-day period when all $^{252}Cf$ data were taken, demonstrating the time spent active by the water in both control sets and $^{252}Cf$ sets. The number in parentheses is the number of nucleations for each data set. The shorter time to nucleation (so, the warmer temperature of nucleation) is evident for Cf (yellow) vs. control (black). Medians are displayed with no fits here of any kind.

The significance of the difference when Cf is compared with itself, control to itself, then Cf compared to control is respectively 1.6σ, 1.5σ, and 4.7σ based on the difference in the medians displayed in Figure 6, divided by all statistical and systematic uncertainties summed in quadrature. As expected, based upon the survival curve differences (Figure 4), source data are self-consistent, control is self-consistent, but $^{252}Cf$ differs from control, significantly, despite being tested under identical conditions except for the source.

It is helpful for potential future applications to identify whether neutrons or gammas interacted with the water molecules/atoms/nuclei. In an attempt at distinguishing neutron- vs. gamma-induced freezing, the Cf data were taken with Pb shielding. With $^{252}$Cf, neutrons are being directly emitted from spontaneous fission; these neutrons are accompanied by a high flux of secondary gammas that are up to several MeV in energy[21]. Varshneya and the theoretical work of others following him including Pisarev's[22] concluded that gammas are more likely to be responsible for water nucleation, not neutrons. Our final analysis we present is thus intended for addressing this possibility, that the gammas not neutrons from the $^{252}$Cf were responsible for the strong effect that we have unequivocally observed. Separated from the Cf by 1 year, the different control data here (Figure 7) are also indicative of the long-term stability of our setup.

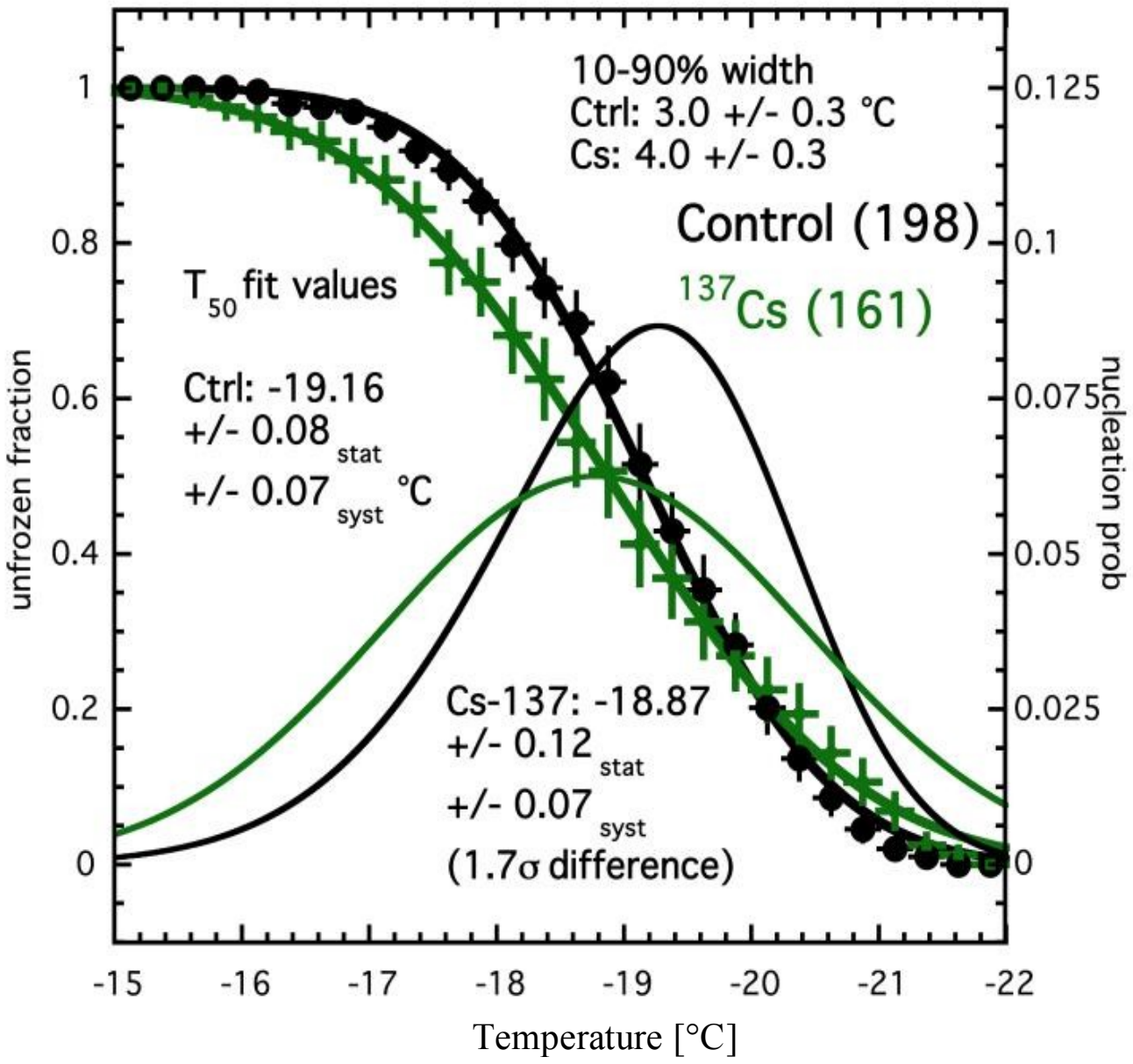

Figure 7. $^{137}$Cs data in green with nearest control black. Skew fits, combined here into a single plot, were chosen as best fitting raw data prior to integration into the 'S' fits (erf) but are mainly visual guides and cross-checks of nucleation probability. As with Cf, skewness is close to 0 for source data. Temperatures (-19.16 and -18.87) cited are $T_{50}$'s, and based on the common-use survival-curve error functions alone.

While Figure 7 is ambiguous, compared with earlier results especially, its very ambiguity is telling: even though the gamma energies (661.66 keV) from the $^{137}$Cs source (10 $\mu$Ci, i.e., 370,000 gammas/second) were lower than from the Cf source, its gamma interaction rate was over two orders of magnitude higher in the water than either the gamma or the neutron interaction rate of the $^{252}$Cf according to full GEANT4 simulations[23] of the setup (G4 is a standard Monte Carlo simulation software tool used for high-energy physics experiments that can account for shielding / geometry). While the difference for the gamma-rays is in the same direction (a higher temperature) as for the Cf it is much less significant, and the Cs curve, based on 161 events, even crosses control's 'S' based on the 198 closest in time (as done for Cf earlier). It is not as disjoint if compared with Figure 4. When the systematic uncertainty is included the statistical significance of the difference from control is merely 1.7σ and can thus not be robustly ruled out as being merely statistical fluctuation. The width is even larger than control, a possible sign of gammas adversely affecting thermocouples directly and causing the temperature measurement to be less reliable[24].

## DISCUSSION

To the best of our knowledge the Cf effect we record has never been reported. While neutrons have been used to study the properties of supercooled water before[25] they have not been used to solidify it as we do here. The 2018 Cf run was most distinctive from the control (source removed) at a maximum of 5.8$\sigma$ in terms of median $T_{min}$ and a minimum of 4.7$\sigma$ in terms of a systematic alternation with control combining the results of both $T_{min}$ and $t_{active}$, in our most conservative analysis. The statistical significance according to the traditional survival-curve analysis was between at 5.2$\sigma$. The gamma-ray data were discrepant from control at 1.7σ only, suggesting neutrons and gammas affect supercooled water disproportionately, with neutrons much more likely to be responsible for increased nucleation probability measured in this work. We nonetheless did explore the possibility that, e.g., gamma-rays from increased n-capture on hydrogen were responsible for the effect we observed in the Cf, but find it not likely, due to the much lower rate compared to other interactions, according to GEANT4.

We note that these results, while significant, are not entirely unexpected. According to both G4 MC sims, as well as simple 2-body 3D collision kinematics, given incoming neutron kinetic energies in the 1 MeV range, the median energy of recoil for hydrogen nuclei was 100 keV from Cf neutrons and 50 keV for oxygen. This is well above the threshold energy described in Khvorostyanov & Curry[37], indicating that nucleation should be expected from the interaction of neutrons in supercooled water, provided sufficient energy is deposited within the critical radius.

In addition, due to the order of magnitude higher stopping power (or differential energy deposition) *dE/dx* from neutrons recoiling against H or O nuclei as opposed to gamma-rays predominantly recoiling from electrons, according to simple Bethe-Bloch calculations, it would be less likely that gamma-rays led to the freezing instead of neutrons, especially if this work with supercooling is analogous to superheating as in bubble chambers[26]. This corroborates our findings.

There is no analysis method presented which leads to a statistical consistency between $^{252}$Cf and control, underscoring the robustness of these results. We acknowledge however that much remains to be studied. The setup needs to be reconstructed with improvements and more data taken with an even greater variety of sources both neutron and gamma, at different distances, with different amounts and types of shielding (Pb for gammas and polyethylene for neutrons). A detailed analysis of scattering cross-sections, and the momentum conservation and kinetic energy transfer required for nucleation at any given temperature is beyond the scope of this initial report, and will be the subject of subsequent publications. An improved Monte Carlo simulation and modelling of the setup is underway for determining the energy threshold, as well as minimum stopping power or linear energy transfer also known as differential energy deposition (or just *dE/dx*) required to trigger a nucleation.

## CONCLUSION

Our paper documents preliminary evidence MeV-scale neutrons cause nucleation of supercooled water. A $5\sigma$ difference from control was measured for a $^{252}$Cf n source. This very promising first result opens up the possibility of employing this technology as an underground experiment which seeks nuclear recoils from the theoretical dark matter particle[28], thought to behave just as a neutron in terms of elastic nuclear scattering[29]. We're encouraged by the initial results that seem to constitute observation of a previously undocumented effect in water, to further develop this technology. If the energy threshold is indeed in the keV regime, this technology could be particularly well-suited to a search for low-mass (sub-GeV/$c^2$ rest-mass energy) dark matter[30]. For that search, the challenges are similar to those faced in measurements of coherent elastic neutrino-nucleus scattering[31]. If there is some sensitivity to $e^-$ recoils, as from gammas, then neutrino-electron interactions in water[32] may also lead to freezing, but this appears to be less likely.

Potential applications of a particle detector based on supercooled water are even more interdisciplinary. It is germane to atmospheric science[33], where conflicting claims of radiation causing nucleation vs. not[1,34] may be resolvable by appeal to different *dE/dx*. Controlling the temperature and/or pressure allows one to control the critical radius, and thus the particle detection thresholds for both energy[35] and *dE/*dx[36], which provides ability to make the detector insensitive (or, selective) to distinctive particle interactions. As the supercooled temperature is lowered, each of the detector response thresholds should be correspondingly lowered. Lower temperatures than reported here should be possible with a greater degree of purification, via use of existing technologies. We coin the term "snowball chamber" to describe the new device, due to spherical ice growth following nucleation(s), the snow-like qualities of the ice produced, and lastly the *snowballing* exothermic reaction whereby an entire volume freezes from only a single initial interaction.

## ACKNOWLEDGMENTS

This work was supported by the University at Albany, State University of New York (SUNY), under new faculty startup funding provided to Prof. Szydagis for 2014-2018, and under the Presidential Innovation Fund for Research and Scholarship (PIFRS) grant awarded to Prof. Szydagis for 2017-18 by the Office of the Vice President for Research, Division of Research, the University at Albany, SUNY, as well as with a FRAP-A award (Faculty Research Award Program 'A') and the startup funds of Prof. Levy. The research was partially supported by the Department of Energy under Award No. DE-SC0015535, allowing R&D part-time. We thank Dr. Aleksey Bolotnikov of Brookhaven National Laboratory for making us aware of previous works, by Pisarev and by Varshneya.

We also gratefully acknowledge the advice, encouragement, and support from the following individuals, who provided their knowledge and their expertise: our UAlbany colleagues Prof. Saj Alam and Prof. Bill Lanford. We thank Dr. Marc Bergevin of Livermore National Lab for helpful discussions on the relative merits of AmBe versus $^{252}$Cf sources. Lastly, Szydagis wishes to personally thank his wife Kel, a linguist not physical chemist by training, for suggesting the filter to use for purification, and coining the name.

## DATA AVAILABILITY

Data in support of the findings of this study are available from the corresponding authors on reasonable request.